\documentclass[twocolumn,prd,superscriptaddress,aps,nofootinbib,floats,floatfix,amsmath,amssymb,secnumarabic]{revtex4}
\usepackage{graphicx,hyperref,verbatim}

\begin{document}

\def\lsim{\mbox{\raisebox{-.6ex}{~$\stackrel{<}{\sim}$~}}}
\def\gsim{\mbox{\raisebox{-.6ex}{~$\stackrel{>}{\sim}$~}}}
\renewcommand{\t}{\tilde}

\title{\bf Initial Conditions for Non-Canonical Inflation}

\author{\normalsize Paul Franche\footnote{franchep@hep.physics.mcgill.ca}}
\affiliation{Department of Physics, McGill University,
3600 University Street, Montr\'eal, Qu\'ebec, Canada H3A 2T8
}
\author{\normalsize Rhiannon Gwyn\footnote{rhiannon.gwyn@kcl.ac.uk}}
\affiliation{Department of Physics, King's College London,
Strand, London, U.K.  WC2R 2LS 
}
\author{\normalsize Bret~Underwood\footnote{bjwood@hep.physics.mcgill.ca}}
\author{\normalsize Alisha Wissanji\footnote{wissanji@hep.physics.mcgill.ca}}
\affiliation{Department of Physics, McGill University,
3600 University Street, Montr\'eal, Qu\'ebec, Canada H3A 2T8
}
\date{\today}

\begin{abstract}
We investigate the dynamics of homogeneous phase space for single-field models of inflation. 
Inflationary trajectories are formally attractors in phase space, but since in practice not all initial conditions 
lead to them,
some degree of fine tuning is required for successful inflation. 
We explore how the dynamics of non-canonical inflation, which has additional kinetic terms that 
are powers of the kinetic energy, can play a role in ameliorating the initial conditions fine tuning problem.
We present a qualitative analysis of inflationary phase space based on the dynamical behavior of the scalar field.
This allows us to construct the flow of trajectories, finding that trajectories generically
decay towards the inflationary solution at a steeper angle
for non-canonical kinetic terms, in comparison to canonical kinetic terms, 
so that a larger fraction of the initial-conditions space leads to inflation.
Thus, non-canonical kinetic terms can be important for removing
the initial conditions fine-tuning problem of some small-field inflation models.
\end{abstract}

\maketitle

\section{Introduction}
\label{sec:intro}

The paradigm of cosmological inflation is an elegant solution to the flatness, horizon, and monopole
problems of standard Big Bang cosmology \cite{Guth}.  The simplest constructions of inflation
consist of a scalar field (the inflaton $\phi$) whose energy density is dominated by potential energy
\cite{ChaoticInflation,NewInflationLinde,NewInflation}.
Remarkably, this simple picture also provides a mechanism for the generation of primordial large scale
perturbations through the quantum fluctuations of the inflaton field.
The spectrum of perturbations is in agreement with cosmological observations for many of the simplest
models (~see e.g. \cite{WMAP,LSS,2dFGRS,Boomerang,DASI}).

However, inflation would lose some of its appeal if special initial conditions are needed for it to occur.
For the purposes of studying inflationary initial conditions, it is useful to divide inflationary models
into two classes based on the distance $\Delta \phi$ the inflaton travels during the inflationary period:
large-field ($\Delta \phi \gg M_p$) inflation and small-field ($\Delta \phi \ll M_p$) inflation \cite{LargeSmallModel}.
The sensitivity of inflation to 
initial conditions was studied extensively
for chaotic inflation \cite{ChaoticInflation} in 
\cite{linde_1985,Belinskyetal,PiranWilliams,Piran:1986dh,GoldwirthPiranReport,brandenberger_kung_1989,
brandenberger_kung_1990,brandenberger_feldman_kung_1991,
goldwirth_piran_1990,goldwirth_1991}, as an example of large-field inflation, 
and for new inflation \cite{NewInflationLinde,NewInflation} in
\cite{Goldwirth,GoldwirthPiranReport,albrecht_brandenberger_matzner_1985,albrecht_brandenberger_matzner_1987,Brandenberger:1988ir,
Brandenberger:1988mc,brandenberger_kung_1989,goldwirth_piran_1990,goldwirth_1991}, as an example of small-field inflation. 
These studies found that there is a noticeable contrast between the two types of models:
the onset of chaotic inflation is insensitive to 
initial conditions, but
the onset of new inflation is very sensitive to initial conditions.  Stated more generally,
small-field models have an initial conditions fine-tuning problem, while large-field models do not.

These studies assumed the dynamics to be that of a single canonical scalar field.  In this paper,
we will take the scalar field to have non-canonical dynamics given by the effective Lagrangian
\begin{equation}
{\mathcal L}_{eff} = p\left(X,\phi\right)\, ,
\label{eq:noncanonL}
\end{equation}
where $X\equiv -\frac{1}{2}(\partial \phi)^2$.
As argued in \cite{NonCanonAttractors,Gelaton}, Lagrangians of this form
can arise as effective field theories by integrating out physics above some intermediate 
energy scale $\Lambda$, including terms that are powers of $X/\Lambda^4$ and $\phi/\Lambda$.
These Lagrangians are interesting to study because they can lead to
inflationary backgrounds which interpolate smoothly between canonical and non-canonical
behavior \cite{NonCanonAttractors}.  The modified kinetic
terms coming from (\ref{eq:noncanonL}) can modify the dynamics of the inflaton in
phase space, potentially changing the sensitivity of small-field inflation to initial conditions.

The inflationary initial conditions problem encompasses both homogeneous and inhomogeneous initial
conditions.  In this paper, we will assume homogeneity in order to focus on the homogeneous
initial conditions problem of small-field inflation.  It is an important question how these
results generalize to the case of inhomogeneous initial conditions.

Non-canonical kinetic terms have been shown to have an important role in the inflationary initial conditions 
fine-tuning problem \cite{AttractiveBrane}, for the specific case of a DBI \cite{DBI} kinetic
term.\footnote{In \cite{Bird:2009pq}, this was investigated in more detail for a specific class of brane inflation models 
\cite{KKLMMT,Delicate,ExplicitDbrane,HolographicSystematics}.  Since DBI inflation does
not occur in a controllable regime, the DBI dynamics do not help to alleviate
the initial conditions fine tuning for these models.}
In this paper, we 
focus on the behavior of the phase space dynamics in more detail for the general class
of non-canonical Lagrangians (\ref{eq:noncanonL}), developing more techniques to understand
the dynamics in the region of phase space far from the inflationary solution.
Other mechanisms for resolving the initial conditions fine-tuning problem include dynamical
fine tuning of the potential \cite{MultibraneFlattening,DynamicalTuning}, tunneling from a false vacuum \cite{Freivogel:2005vv},
time-dependent corrections to the potential \cite{Itzhaki1},
moduli trapping at enhanced symmetry points \cite{Beauty},
and different choices of measure on the space of initial conditions \cite{NaturalMeasure,MeasureCosmo}.

In Section \ref{sec:review} we review the scenario of non-canonical inflation with a Lagrangian of the form (\ref{eq:noncanonL}).
The general organizational scheme of the rest of the paper is to examine the phase space and 
dynamics for canonical and non-canonical inflation
in increasing levels of detail.  
First, in Section \ref{sec:structure}, we 
divide phase space into several dynamical regions that allow us to make general statements about
phase-space flow.  
In Section \ref{sec:angles} we analytically compute the angle trajectories 
make in phase space, relative to the horizontal, finding that non-canonical kinetic terms make individual trajectories
steeper towards the inflationary attractor.  
Next, in Section \ref{sec:overshoot}, we focus on the region of phase space with large momentum where most 
overshooting initial conditions are located.
We show that non-canonical kinetic terms can modify the dynamics in this region so that these
initial conditions no longer lead to overshooting.
In Section \ref{sec:examples} we study the phase space of specific examples numerically, confirming
that canonical small-field models have an initial conditions fine-tuning problem while corresponding
non-canonical models do not.  
In particular, we explicitly compute the size of initial-conditions phase space leading to more than
60 e-folds of inflation for both canonical and non-canonical Lagrangians and show that the latter
is much larger than the former.
We conclude with some closing discussion in Section \ref{sec:conclusion}.  
Computations for an additional non-canonical Lagrangian, the power-like Lagrangian, 
not included in the main text are presented in the Appendix.

\section{Review of Non-Canonical Inflation}
\label{sec:review}

Non-canonical inflation has a generalized scalar field Lagrangian minimally coupled\footnote{A non-minimal
coupling can be written in this form after an appropriate Weyl transformation to Einstein frame \cite{Easson,Easson2}.} to gravity,
\begin{equation}
S = \int d^4x \sqrt{g_4}\left[\frac{M_p^2}{2}  {\mathcal R}_4 + p(X,\phi)\right]\, ,
\label{eq:Action}
\end{equation}
where the Lagrangian is a function of 
$X\equiv -\frac{1}{2} (\partial_\mu \phi)^2$ and $\phi$ only.\footnote{As was argued in 
\cite{NonCanonAttractors}, higher-derivative terms in the scalar
field action can be ignored for constructing background non-canonical inflationary solutions.}

In this paper, we will be interested in {\it homogeneous} cosmological backgrounds of the form
\begin{equation}
ds^2 = -dt^2 + a(t)^2 d\vec{x}^2\, ,\, \, \, \,\,\,\,\,\,\,\, \phi = \phi(t)\, ,
\end{equation}
so that $X= \frac{1}{2} \dot{\phi}^2 > 0$.
Non-canonical inflation is described by the inflationary parameters \cite{NonGauss}
\begin{eqnarray}
&& H \equiv \frac{\dot a}{a};\ \ \ \epsilon \equiv -\frac{\dot H}{H^2};\ \ \ \eta \equiv \frac{\dot \epsilon}{H \epsilon};\ \ \ 
	\kappa \equiv \frac{\dot c_s}{H c_s}; \nonumber \\
&& c_s^2 \equiv \left(1+2 X \frac{\partial^2 p \partial_X^2}{\partial p/\partial X}\right)^{-1}\, .\nonumber
\end{eqnarray}
The parameters $\epsilon,\eta$ are generalizations of the usual slow-roll parameters to a general cosmological
background, and the sound speed $c_s$ is the speed at which scalar perturbations travel.
The phenomenology of non-canonical inflation, including the scalar power spectrum $P_k^{\zeta}$, 
the scalar spectral index $n_s$, the tensor-to-scalar ratio $r$, and the (equilateral)
non-gaussianity $f_{NL}^{(equil)}$ is \cite{kinflationPert,NonGauss}
\begin{eqnarray}
P_k^{\zeta} &= & \frac{1}{8\pi^2}\frac{H^2}{M_p^2} \left.\frac{1}{c_s \epsilon}\right|_{c_s k = a H}\,  ;\\
n_s -1 &=& -2\epsilon-\eta-\kappa \, ;\\
r &=& 16 c_s\, \epsilon \, ; \\
f_{NL}^{(equil)} &\sim & c_s^{-2}\, .
\end{eqnarray}

Not all Lagrangians $p(X,\phi)$ are suitable as physically realistic field theories.  In particular,
we would like our scalar field to be ghost free with manifestly subluminal speed of perturbations
(see however \cite{Babichev:2007dw} for non-subluminal cases).
These conditions together require \cite{SickTheories,Bean}
\begin{equation}
\frac{\partial p}{\partial X} > 0\,\,\,\, \mbox{ and } \,\, \frac{\partial^2 p}{\partial X^2} > 0\, .
\label{eq:physical}
\end{equation}
For simplicity, we will focus on Lagrangians that have non-zero potential energy and take a ``separable"
form of potential and kinetic terms,
\begin{equation}
p(X,\phi) = q(X,\phi) - V(\phi)\,.
\label{eq:separable}
\end{equation}
We will constrain the kinetic term to obey the physicality constraints (\ref{eq:physical}) and to
reduce to a canonically normalized kinetic term 
$q(X,\phi) \approx X$ for sufficiently small $X$.

The homogeneous scalar field can be treated as a perfect fluid with pressure $p = p(X,\phi)$
and, under the assumption (\ref{eq:separable}), the energy density separates nicely
into kinetic and potential energies:
\begin{equation}
\rho \equiv 2X \frac{\partial p}{\partial X} - p = 2X \frac{\partial q}{\partial X}-q + V(\phi) \, .
\end{equation}
The Friedmann equation for the scale factor then becomes
\begin{equation}
H^2 = \frac{1}{3 M_p^2} \left[2X \frac{\partial q}{\partial X} - q + V(\phi)\right]\, .
\label{eq:Hubble}
\end{equation}
The equation of motion for the scalar field can be written compactly in terms of the canonical
momentum
\begin{eqnarray}
\label{eq:Pidef}
\Pi &\equiv & -\frac{\partial p}{\partial \dot \phi} = -\dot \phi \frac{\partial p}{\partial X} = \sqrt{2X} \frac{\partial q}{\partial X}; \\
\dot \Pi &=& -3 H \Pi - \frac{\partial p}{\partial \phi}\, .
\label{eq:PiEOM1}
\end{eqnarray}
In defining $\Pi$ we assumed $\dot \phi < 0$, which is the typical case we will consider below,
and added a minus sign to the usual definition of $\Pi$.  We have chosen the latter convention because
we will prefer to work in only one quadrant of phase space where $\phi \geq 0$ and $\dot \phi \leq 0$;
the additional minus sign allows us to work in the purely positive quadrant of phase space
$\phi \geq 0$ and $\Pi \geq 0$, where
there will be no more ambiguities due to minus signs.

For canonical kinetic terms, slow-roll inflation occurs when the potential energy dominates over the
kinetic energy, i.e.  $\frac{1}{2}\dot \phi^2 \ll V(\phi)$, and the acceleration is small compared to the Hubble friction
and driving force, that is $\ddot \phi \ll 3H\dot\phi$ and $\ddot \phi \ll V'$ , leading to the 
slow-roll inflationary solutions $\dot\phi_{slow-roll} = -V'/3H$.
Non-canonical inflation is a straightforward generalization of this \cite{NonCanonAttractors}, with
the potential energy dominating the kinetic energy, i.e. 
\begin{equation}
V(\phi) \gg 2X \frac{\partial q}{\partial X} - q,
\end{equation}
and the Hubble friction and driving 
force\footnote{In general, non-canonical inflationary solutions occur when the 
driving force term $\partial p/\partial \phi$
of the scalar field equation of motion (\ref{eq:PiEOM1}) is dominated by the driving force from the potential energy,
so that $\partial p/\partial \phi \approx -V'$ \cite{NonCanonAttractors}.}
dominating over the ``acceleration":
\begin{equation}
\dot \Pi \ll 3H \Pi, V'\,.
\end{equation}
The corresponding non-canonical inflationary solutions are then
\begin{equation}
\Pi_{inf}(\phi) = \frac{V'}{3H}\, .
\label{eq:PiInf}
\end{equation}

We will focus on Lagrangians that can be written as a power series,
\begin{equation}
p(X,\phi) = \sum c_n(\phi) \frac{X^{n+1}}{\Lambda^{4n}} - V(\phi),
\end{equation}
which converges in some finite radius of convergence, i.e. $X/\Lambda^4 \in [0,R]$, for $0< R < 1$.  As discussed in \cite{NonCanonAttractors}, sufficient conditions for non-canonical inflation to occur for these types
of Lagrangians are 
(a) the derivative of the series with 
respect to $X$, $\partial_X p$, must diverge at the boundary of the radius
of convergence; 
(b) the dimensionless parameter 
\begin{equation}
A \equiv V'/(3H\Lambda^2)
\end{equation}
must be large, $A \gg 1$; and 
(c) the coefficients
$c_n$ must satisfy $\Lambda^2 \partial_\phi c_n/(3H c_n) \ll 1$.  For simplicity, we will choose the coefficients $c_n$
to be constant throughout this paper so that the latter condition is always satisfied.\footnote{It would be interesting
to consider how generic this is from an effective field theory point of view.} 
Condition (b) can be rewritten in terms of the {\it slow-roll} parameter
\begin{equation}
\epsilon_{SR} \equiv \frac{M_p^2}{2} \left(\frac{V'}{V}\right)^2\, 
\label{eq:SRepsilon}
\end{equation}
as $A = \left(\frac{2}{3} \epsilon_{SR} \frac{V}{\Lambda^4}\right)^{1/2}$.
As discussed in \cite{NonCanonAttractors}, in order to have $A \gg 1$ we need $V/\Lambda^4 \gg 1$.
Typical examples of Lagrangians that satisfy the condition that the derivative of the series diverge
at the boundary of the radius of convergence are the DBI Lagrangian \cite{DBI}
and the ``geometric series" Lagrangian,
\begin{eqnarray}
p(X,\phi)_{DBI} &=& -\Lambda^4 \left(\sqrt{1-2\frac{X}{\Lambda^4}}-1\right) - V(\phi)\, ;\\
p(X,\phi)_{geo} &=& \Lambda^4 \left(\frac{1}{1-\frac{X}{\Lambda^4}}-1\right)-V(\phi)\,, 
\end{eqnarray}
with $R=\{1/2,1\}$, respectively.  
For $X/\Lambda^4 \ll 1$ these Lagrangians consist of a canonical kinetic term plus subleading corrections,
while for $X \rightarrow R \Lambda^4$ the kinetic terms are far from canonical.

\section{General Phase Space Structure}
\label{sec:structure}

Before we discuss in detail the dynamics of inflationary models in $(\phi,\Pi)$ 
phase space it is helpful to first determine the general structure of this phase space.

\begin{figure}[tp]
\includegraphics[scale=.43]{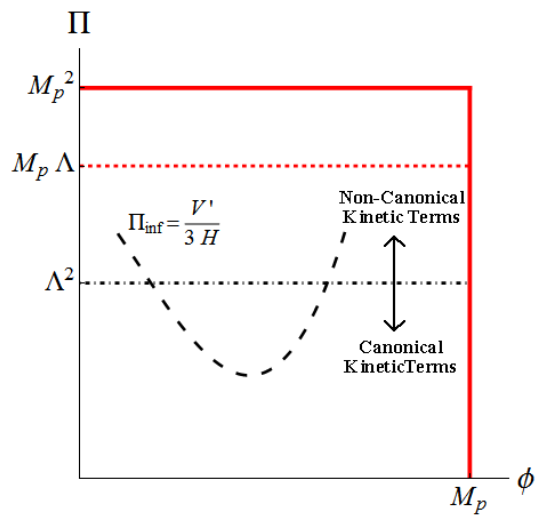}
\caption{\small Homogeneous phase space is bounded by $H<\Lambda$, resulting in the solid (red) lines
for non-canonical Lagrangians and the dashed line for a canonical scalar field.  The dot-dashed line
corresponds to the boundary between canonical and non-canonical behavior of the kinetic terms.}
\label{fig:AvailablePhaseSpace}
\end{figure}

First, we note that the inflationary solution (\ref{eq:PiInf})
corresponds to a one-dimensional trajectory in phase space, as shown in Figure \ref{fig:AvailablePhaseSpace}.  
Since the inflationary solution is only valid when the
inflationary parameters are small, the trajectory does not extend
throughout all of phase space.  In \cite{NonCanonAttractors} it was argued that the dimensionless
parameter $A$ (defined in the previous section) keeps track of whether the inflationary solution is predominately
canonical (slow roll) or non-canonical.  When $A \ll 1$, we have the usual slow roll inflation, while for
$A \gg 1$ inflation is strongly non-canonical.
Graphically, this corresponds to the inflationary trajectory
being below, or above, the line $\Pi = \Lambda^2$ in phase space, respectively.  More generally, the non-canonical
kinetic terms are important when $X/\Lambda^4\rightarrow R \sim {\mathcal O}(1)$.
Since $q' = \partial q/\partial X > 1$ always, we see
that the region of phase space where the kinetic terms are non-canonical corresponds to the region with
\begin{equation}
\frac{\Pi}{\Lambda^2} = \sqrt{2\frac{X}{\Lambda^4}} q'\sim \sqrt{2R}\, q' \gsim 1\, .
\end{equation}
Correspondingly, the region of phase space below this line is the canonical kinetic term regime.

Clearly, the available phase space must be bounded - our effective description in terms
of a single scalar field must break down at the very least for large momenta and potential energies.
One way the effective description breaks down is when the Hubble rate of the background exceeds
the scale of new physics, i.e. $H > \Lambda$.  When this occurs, the high-energy physics can no longer be
integrated out and must be included in the dynamics.
To derive bounds on the momentum, assume that the energy density is dominated by the kinetic energy
(which is true for large momentum):
\begin{eqnarray}
H^2 &&= \frac{1}{3M_p^2} \left[2X q' - q + V\right] > \frac{2X q'-q}{3M_p^2} \nonumber \\
&&\sim \frac{2X q'}{3M_p^2} = \frac{\sqrt{2X}\, \Pi}{3 M_p^2}\, ,
\end{eqnarray}
where we used the definition of the momentum (\ref{eq:Pidef}).

In the case of a canonical kinetic term, we have $\Pi = \sqrt{2X}$ and the 
bound $H < \Lambda$ turns into $\Pi < M_p \Lambda$
(this is conservative; if we had chosen the usual $H<M_p$ the initial
conditions fine-tuning problem for canonical models would just be worse).  
The kinetic energy of Lagrangians that lead to non-canonical inflation, $\rho_{kin} \sim \sqrt{2X}\,\Pi$, tends to grow
{\it slower} with momentum $\Pi$ than a canonical kinetic energy, $\rho_{kin}^{canon} \sim \Pi^2$.
For instance, when the higher powers in $X$ are important and $X/\Lambda^4 \sim R$, 
power series Lagrangians have kinetic energies that scale
linearly with momentum $\rho_{kin}^{power-series} \sim \sqrt{2R}\, \Pi$.  
Because the kinetic energy grows slower in $\Pi$,
the bound on the momentum from the effective field theory constraint $H < \Lambda$ will
be more relaxed than in the canonical case; $\Pi < 3(2R)^{-1/2} M_p^2$.
(This is different from the case in \cite{AttractiveBrane,Bird:2009pq} because
these works took the kinetic term to be canonical in deriving the bound.)
There is a large range of large momenta $\Lambda^2 < \Pi \lsim M_p^2$ for which
the non-canonical kinetic terms are important in phase space and the effective field theory treatment
of the background is still (naively) valid.

There are additional constraints on the validity of the effective field theory that arise from considering
the strength of the couplings of the perturbations generated during inflation 
\cite{EFTInflation,PerturbativeInflation,Shandera:2008ai}.  If these perturbations are too strongly coupled,
they can backreact on the (homogeneous) inflating background so that the perturbation spectrum and background
evolution are untrustable.  We will work in the regime where these constraints are satisfied;
when we consider specific examples, in Section \ref{sec:examples}, we will choose our parameters such
that these constraints are satisfied thoughout the regime of interest.  We refer the reader to \cite{NonCanonAttractors}
for a more detailed discussion of these effective field theory constraints.

In principle, the bound $H < \Lambda$ also places restrictions on the allowed range of the 
scalar field $\phi$ through the potential energy.  In practice, other difficulties arise
when $\phi$ is large, such as $\phi$-dependent masses or couplings for other fields becoming
important.  We are primarily concerned here with inflationary potentials which, for canonical kinetic
terms, suffer from an initial conditions fine-tuning problem.  As discussed in the introduction, 
these potentials are almost always of the {\it small-field} type,
so that typical values of the scalar field are small in Planck units $\phi \ll M_p$.
We will take $\phi=M_p$ as an upper bound on the field range so that we are always working in the
small-field regime.  Specific models may have additional
stronger bounds on the field range that should be satisfied
(such as D-brane inflation \cite{BaumannMcAllister}), but we will try to be as general as possible here and not restrict
ourselves to any one particular model.

\begin{figure*}[tp]
\centerline{
\includegraphics[scale=.5]{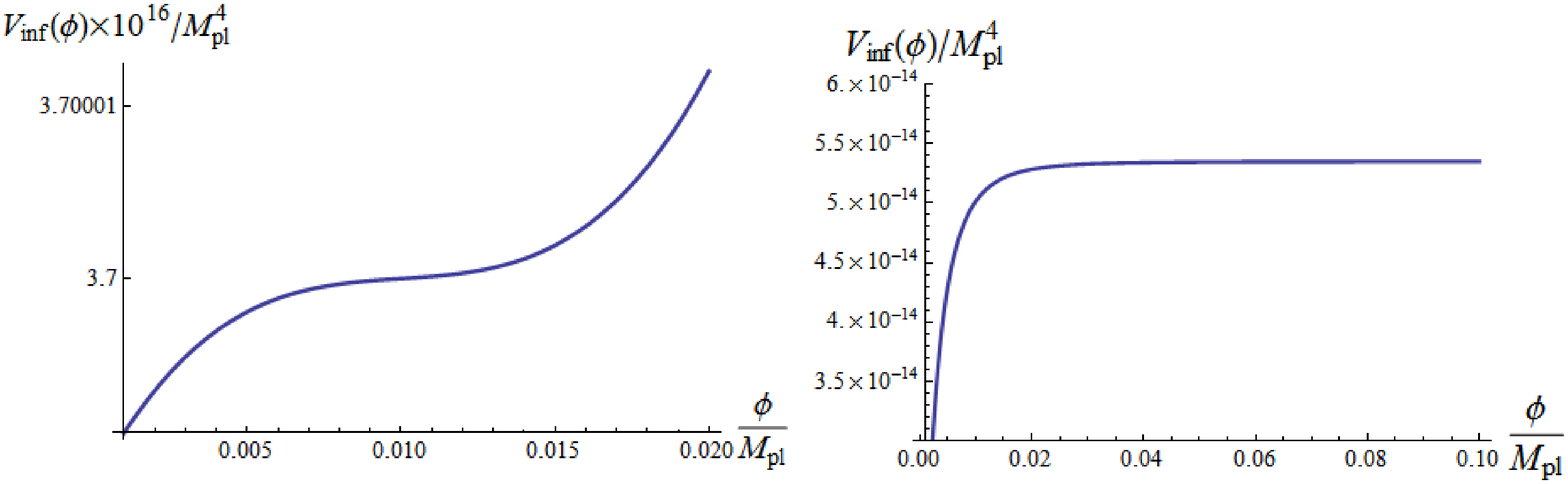}}
\caption{\small Left: An inflection point type potential (\ref{eq:inflectionpt}).  Right:
A coulomb-type potential (\ref{eq:coulomb}).}
\label{fig:potentials}
\end{figure*}

\begin{figure*}
\includegraphics[scale=.52]{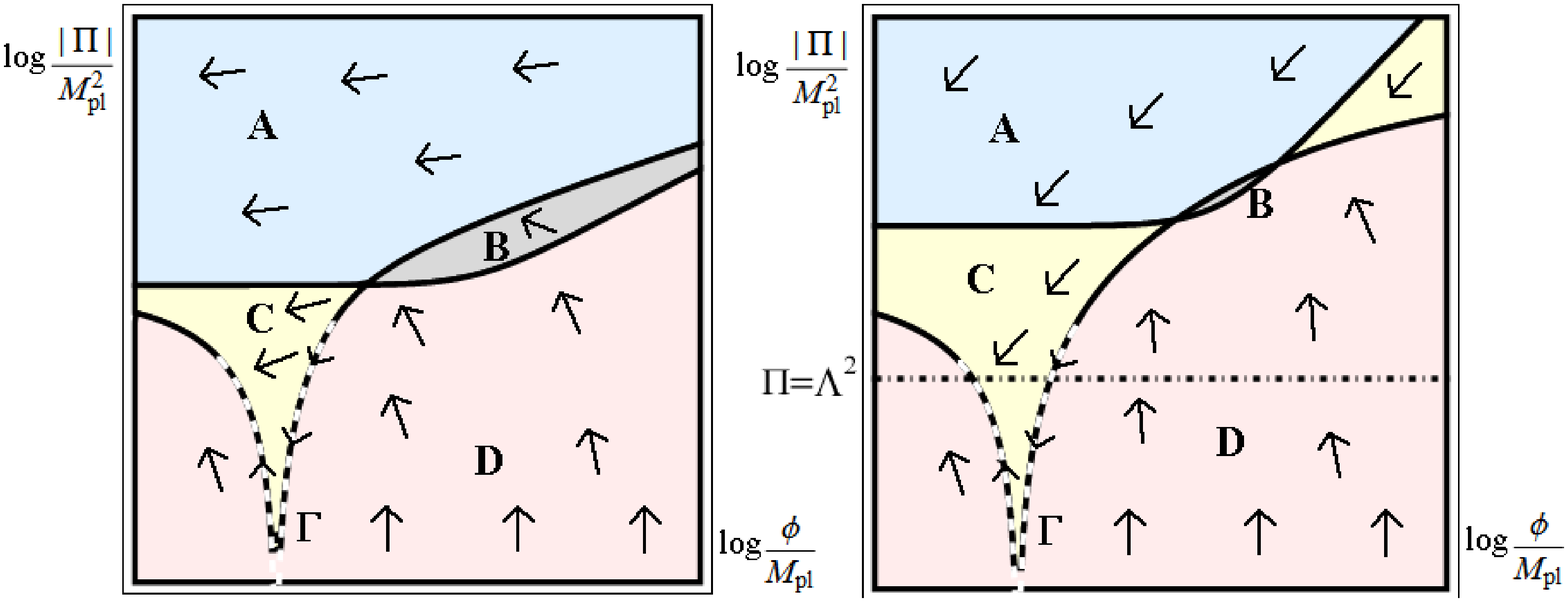} 
\caption{\small A sketch of the boundaries for the Regions A-D of $(\phi,\Pi)$ phase space described in the text for an inflection-point type potential (\ref{eq:inflectionpt}) with canonical (left) and non-canonical (right) kinetic terms.
The inflationary solution $\Pi_{inf}(\phi)$ (\ref{eq:PiInf}) is the dashed
curve $\Gamma$.  The horizontal dot-dashed line is $\Pi = \Lambda^2$; above this line the non-canonical kinetic term is very
far from canonical.  Notice how Regions A and B shrink when going to the non-canonical case, as discussed in the text.
Arrows denote the general direction of flow of trajectories, and are generally steeper for non-canonical kinetic terms.}
\label{fig:RegionPlots_Inflection}
\end{figure*}

\begin{figure*}
\includegraphics[scale=.5]{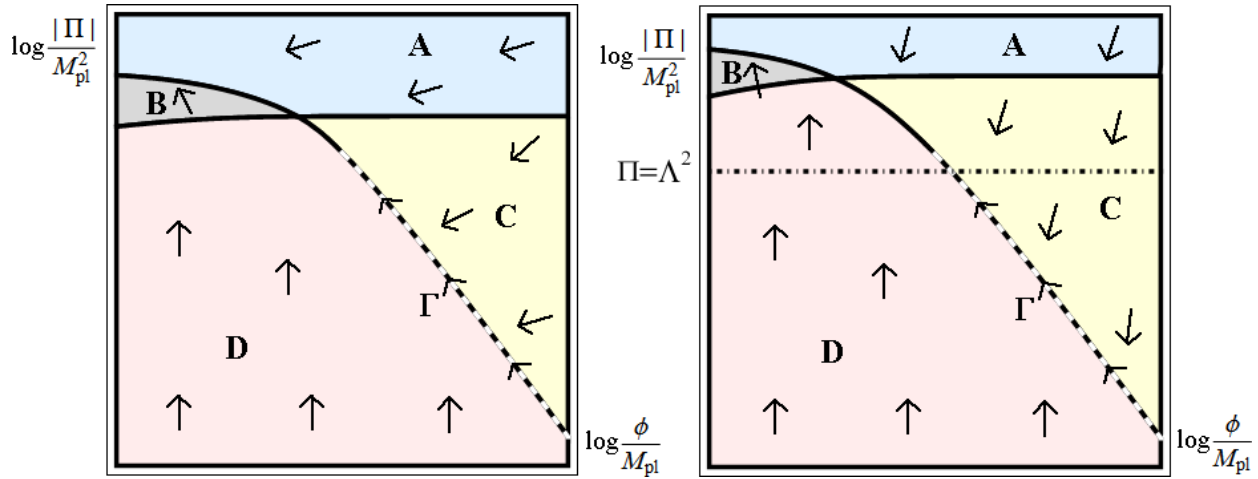}
\caption{\small The boundaries for the Regions A-D of $(\phi,\Pi)$ phase space described in the text for a 
coulomb-type potential (\ref{eq:coulomb}) with canonical (left) and non-canonical (right) kinetic terms.
The differences between the canonical and non-canonical cases are similar to those discussed in Figure 
\ref{fig:RegionPlots_Inflection}.}
\label{fig:RegionPlots_Coulomb}
\end{figure*}

The $(\phi,\Pi)$ phase space\footnote{Note that here we will not be using the full $(\phi, \Pi, H)$ phase space as in
\cite{Felderetal}, since $H$ is simply set by a constraint equation and is not separately dynamical.}
has a general structure, based on the equations of motion,
\begin{eqnarray}
\label{eq:PiEOM}
\dot \Pi &=& -3 H \Pi + V'; \\
H^2 &=& \frac{1}{3M_p^2}\left[2X\frac{\partial q}{\partial X} -q + V(\phi)\right]\, ,
\label{eq:HEOM}
\end{eqnarray}
from which the general qualitative dynamics in phase space can be extracted.
In particular, we can divide phase space into several regions of
interest, depending on the type of terms that dominate the equations of motion (\ref{eq:PiEOM},\ref{eq:HEOM})
(see Figures \ref{fig:RegionPlots_Inflection} and \ref{fig:RegionPlots_Coulomb}):
\begin{itemize}

\item {\bf Region A:}
Hubble friction dominates the driving force from the slope of the potential, 
$3H \Pi \gg V'$, and kinetic energy dominates
over potential energy $\left(2 X \frac{\partial q}{\partial X} - q\right) \gg V(\phi)$.  These conditions
are satisfied in the large $\Pi$ region of phase space.
Clearly in this region we have $\dot \Pi < 0$, so momentum {\it decreases} with time.

\item {\bf Region B:} 
The driving force of the potential dominates over the Hubble friction,  $3H \Pi \ll V'$, but kinetic energy dominates over potential energy; 
$\left(2 X \frac{\partial q}{\partial X} - q\right) \gg V(\phi)$.  These conditions are met
when the potential has a large slope.  In this region, we have $\dot \Pi > 0$, so momentum
{\it increases} with time.

\item {\bf Region C:} 
Hubble friction dominates, but potential energy dominates over
kinetic energy.  Trajectories in this region lose momentum since $\dot \Pi < 0$; the lower boundary 
of this region contains the inflationary trajectory.

\item {\bf Region D:} 
In this remaining region, the potential energy driving force dominates
over the Hubble friction and the energy density is dominated by the potential energy.  Trajectories
in this region gain momentum $\dot \Pi > 0$; the upper boundary of this region contains the inflationary
trajectory.

\end{itemize}

A similar regional analysis appears in \cite{Goldwirth,GoldwirthPiranReport}
for the new inflation and chaotic inflation 
models (see also \cite{Brandenberger:2003py}).
In this paper we will be interested instead in small-field inflationary models with
potentials of two types that are meant to be representative of many of the potentials that
arise from modern top-down inflationary model-building. The inflection-point type potential,
\begin{equation}
\label{eq:inflectionpt}
V(\phi)_{inflection} = V_0 + \lambda (\phi-\phi_0) + \beta (\phi-\phi_0)^3, \\
\end{equation}
arises in models of D-brane inflation \cite{Delicate,ExplicitDbrane,HolographicSystematics} and
specific closed string inflation models \cite{Accidental,Badziak:2008gv}.
The coulomb-type potential,
\begin{equation}
V(\phi)_{coulomb} = V_0 - \frac{T}{(\phi+\phi_0)^n}\, ,
\label{eq:coulomb}
\end{equation}
has appeared as a potential for D-brane inflation
\cite{DvaliTye} and its embeddings in warped geometries \cite{KKLMMT,HolographicSystematics}.
These potentials are shown in Figure \ref{fig:potentials}.

It is helpful to divide phase space into these regions because it is straightforward to determine
the broad structure of the flow of trajectories in each region without solving the numerical equations
of motion.  In particular, we know that in Regions A and C trajectories tend to lose momentum,
while in Regions B and D trajectories tend to gain momentum.  
Clearly, the boundary between Regions C and D contains the inflationary trajectory, which we will
label as $\Gamma$.
Note that $\Gamma$ does not encompass the entire boundary between Regions C and D, but rather only
a fraction of it.  This is because $\Gamma$ is defined only when $\epsilon, |\eta| \ll 1$, while the boundary
between Regions C and D exists more generally.
The dividing line $\Pi = \Lambda^2$ between the canonical and non-canonical regimes can cut through
anywhere on this region plot.  For models in which some amount of non-canonical inflation occurs, the dividing line
must cut through some part of the inflationary trajectory $\Gamma$.

It is straightforward to sketch the boundaries for the Regions by finding the inequalities that determine whether a)
the energy density is kinetic or potential energy dominated, and b) the $\Pi$ equation of motion (\ref{eq:PiEOM})
is Hubble friction or driving force dominated, for both the canonical and non-canonical limits.
For a canonical kinetic term, the energy density is
\begin{equation}
\rho = \frac{1}{2}\Pi^2 + V(\phi),
\end{equation}
so the condition that the kinetic energy dominates is
\begin{equation}
\Pi > \sqrt{2V}\, .
\label{eq:canonKE}
\end{equation}
For a non-canonical kinetic term the energy density is
\begin{equation}
\rho = 2X q'(X) - q(X) + V(\phi)\sim \sqrt{2X} \Pi + V(\phi)
\end{equation}
so in the non-canonical limit ($X\rightarrow \Lambda^4 R$) the condition that the kinetic energy dominates
becomes
\begin{equation}
\Pi > \frac{V}{\Lambda^2 \sqrt{2R}} = \frac{\sqrt{2V}}{\sqrt{2R}} \left(\frac{V}{2\Lambda^4}\right)^{1/2}\, .
\label{eq:noncanonKE}
\end{equation}
Recalling that non-canonical inflation requies $V/\Lambda^4 \gg 1$ \cite{NonCanonAttractors}, we see that the condition (\ref{eq:noncanonKE})
is stronger than the condition (\ref{eq:canonKE}).  Thus, we expect that the region where the kinetic energy
dominates will shift to larger momentum for non-canonical kinetic terms.

Now we consider the conditions necessary for Hubble friction to dominate the $\Pi$ equation of motion (\ref{eq:PiEOM}).
This itself can occur in two subcases: the Hubble friction can dominate while the energy density is dominated by
either kinetic or potential energy, corresponding to the boundaries between Regions A and B or C and D, respectively.
When the kinetic energy dominates, for a canonical kinetic term the condition that the Hubble friction dominates can
be written in terms of the {\it slow-roll} parameter (\ref{eq:SRepsilon}), as
\begin{eqnarray}
\Pi &>& \sqrt{2V} \left(\frac{\epsilon_{SR}}{3}\right)^{1/4}\, .
\label{eq:canonHubbleKE}
\end{eqnarray}
Notice that this implies that Region B, which corresponds to kinetic energy and driving force domination, can
only occur (in the canonical limit) for potentials that have $\epsilon_{SR} > 3$.\footnote{A more precise analysis, 
not assuming the kinetic energy dominates, finds that the condition becomes $\epsilon_{SR} > 6$.}
In the non-canonical limit, these conditions become
\begin{equation}
\Pi > \left(\frac{\epsilon_{SR}^2}{18}\right)^{1/6} \frac{\sqrt{2V}}{(2R)^{1/6}}\left(\frac{V}{\Lambda^4}\right)^{1/6}\, .
\label{eq:noncanonHubbleKE}
\end{equation}
Comparing (\ref{eq:noncanonKE}) and (\ref{eq:noncanonHubbleKE}), we see that the conditions for Region B become even more difficult
to satisfy, i.e. $\epsilon_{SR} > \frac{3V}{4R \Lambda^4}$.
Now, for the case when the Hubble friction and potential energy dominate, the Hubble friction bound is the same for both canonical and
non-canonical kinetic terms (because the $\Pi$ equation of motion is the same for both):
\begin{equation}
\Pi > \sqrt{2V} \left(\frac{\epsilon_{SR}}{3}\right)^{1/2}\, .
\label{eq:HubblePE}
\end{equation}

Combining these bounds (\ref{eq:canonKE}-\ref{eq:HubblePE}) allows us to sketch the boundaries of the Regions A-D.
In Figures \ref{fig:RegionPlots_Inflection}, and \ref{fig:RegionPlots_Coulomb}
these regions are sketched for both canonical and non-canonical kinetic terms with the inflection point and
coulomb potentials, where
we see the qualitative features discussed above.

\section{Angles in Phase Space}
\label{sec:angles}

In the previous section we saw how phase space has a general structure that can be useful for
qualitatively understanding the dynamics of different initial conditions.
In particular, our regional analysis allows us to map out where in phase space we expect trajectories to be
gaining or losing momentum.  However, this analysis is not enough to make judgements about the sensitivity
(or lack thereof) of inflation to initial conditions because we do not know the {\it angle} the trajectory
makes in phase space.  Let us define the angle of the trajectory in phase space as in Figure \ref{fig:AngleDiagram}
(where the factor of $M_p$ is needed to make the angle dimensionless):
\begin{equation}
\tan \theta = \frac{1}{M_p}\frac{d\Pi}{d\phi} = \frac{1}{M_p}\frac{\dot \Pi}{\dot \phi} \, .
\label{eq:angledef}
\end{equation}
All we know from the regional analysis of the previous section is that $\theta > 0$ in Regions A and C
and $\theta< 0$ in Regions B and D.  However, trajectories in Region A with $\theta \ll 1$ (i.e. nearly horizontal)
lead to a very different qualitative picture of phase space than do trajectories with $\theta \sim \frac{\pi}{2}$ (i.e. nearly vertical),
so not only the sign but also the {\it magnitude} of the angle is important.

For a canonical kinetic term
the angle can be computed using the equation of motion (\ref{eq:PiEOM}) and the definition of the momentum
($\Pi = -\dot \phi$):
\begin{equation}
\tan \theta_{canon} = -\frac{\dot \Pi}{\Pi M_p} = \frac{3H\Pi - V'}{\Pi M_p}\, .
\label{eq:canonangledef}
\end{equation}
It is straightforward to evaluate this angle for the various regions, and we find
\begin{equation}
\tan \theta_{canon} = \begin{cases}
\sqrt{\frac{3}{2}} \frac{\Pi}{M_p^2} & \mbox{Region A} \cr
-\frac{\sqrt{2\epsilon_{SR}} V}{\Pi M_p^2} & \mbox{Region B} \cr
\sqrt{3}\frac{\sqrt{V}}{M_p^2} & \mbox{Region C} \cr
-\frac{\sqrt{2\epsilon_{SR}} V}{\Pi M_p^2} & \mbox{Region D.} \cr
\end{cases}
\end{equation}
Since $\Pi \ll M_p$, we see that in Region A $\theta_{canon} \ll 1$, so {\it trajectories are approximately horizontal} here.
These horizontal trajectories are overshoot trajectories.

In Region B we have $\sqrt{2V} < \Pi < \sqrt{2V} (\epsilon_{SR}/3)^{1/4}$ so the size of the angle in this region is bounded by
\begin{equation}
(3 \epsilon_{SR})^{1/4} \frac{\sqrt{V}}{M_p^2} < |\tan \theta_{canon}| < \epsilon_{SR}^{1/2} \frac{\sqrt{V}}{M_p^2}\, .
\nonumber
\end{equation}
Because $V \ll M_p^4$, in Region B we expect $|\theta_{canon}| \lsim 1$ (recall $\epsilon_{SR} > 3$ in Region B).
Similarly, in Region C the angle is small, i.e. $\theta_{canon} \ll 1$.
In Region D, clearly $\theta_{canon} \rightarrow -\pi/2$ as $\Pi \rightarrow 0$,  so trajectories
in Region D are mostly vertical.
Certainly, by the definition of the trajectory angle (\ref{eq:canonangledef}), we have $\theta \approx 0$ along the inflationary
solution (because $\tan \theta = \dot \Pi/(\Pi M_p) \sim (H/M_p) {\mathcal O}(\epsilon)$, as discussed in \cite{NonCanonAttractors},
the angle is not exactly horizontal along the inflationary trajectory).
The qualitative features of these angles are represented in the left-hand plots of 
Figures \ref{fig:RegionPlots_Inflection} and \ref{fig:RegionPlots_Coulomb}.

Now let us assume that we have some non-canonical kinetic terms that become important for $\Pi > \Lambda^2$.
For simplicity, take the scale $\Lambda$ such that the transition from canonical to non-canonical cuts through
Regions C and D as in Figures \ref{fig:RegionPlots_Inflection} and \ref{fig:RegionPlots_Coulomb}, so that Regions A and B are entirely
in the non-canonical regime.

Again, using the equation of motion and the definition of the momentum, the trajectory angle
in the limit of non-canonical kinetic terms (where $X \rightarrow R \Lambda^4$) is
\begin{equation}
\tan \theta_{non-canon} = -\frac{\dot \Pi}{M_p \sqrt{2X}} = \frac{3H\Pi - V'}{\sqrt{2R} M_p \Lambda^2}\, .
\end{equation}
The angles for the different regions in the non-canonical case are then
\begin{equation}
\tan \theta_{non-canon} = \begin{cases}
\frac{\sqrt{3}}{(2R)^{1/4}} \left(\frac{\Pi}{M_p^2}\right)^{3/2} \left(\frac{M_p}{\Lambda}\right) & \mbox{(A)} ;\cr
-\frac{\sqrt{2\epsilon_{SR}} V}{\sqrt{2R} M_p^2 \Lambda^2} & \mbox{(B) and (D)}; \cr
\frac{\sqrt{3 V} \Pi}{\sqrt{2R} M_p^2 \Lambda^2} & \mbox{(C)}\, . \cr
\end{cases}
\end{equation}
Comparing the angle for Region A to the canonical case, we see that the non-canonical angle is larger by a factor of
\begin{equation}
\frac{\tan \theta_{non-canon}}{\tan \theta_{canon}} = \frac{\sqrt{2}}{(2R)^{1/4}} \left(\frac{\Pi}{M_p^2}\right)^{1/2} 
	\left(\frac{M_p}{\Lambda}\right) \gg 1,
\label{eq:thetacompA}
\end{equation}
which is larger than one because $\Pi \gg \Lambda^2$ in Region A.  Thus, non-canonical trajectories in Region A are steeper
than their canonical counterparts, an important fact to note for the overshoot problem.
A similar comparison for Region B and the parts of Regions C and D that are non-canonical leads to
\begin{equation}
\frac{\tan \theta_{non-canon}}{\tan \theta_{canon}} = \frac{\Pi}{\sqrt{2R}\Lambda^2} \gg 1,
\end{equation}
so that, quite generally, {\it non-canonical trajectories are always steeper} than their canonical counterparts.
Of course, the angles in the small $\Pi$ parts of Regions C and D are the same as in the canonical case
discussed above, because the kinetic term is canonical there.

Another way to see how trajectories with non-canonical kinetic terms are generically steeper 
is to notice that the angle for non-canonical kinetic terms in 
Regions B to D is just a multiple of the angle for canonical kinetic terms,
\begin{eqnarray}
\tan \theta_{non-canon} &=& \frac{1}{M_p}\frac{\dot \Pi}{\dot \phi} = \frac{\partial p}{\partial X} \left(\frac{3H\Pi - V'}{\Pi M_p}\right)\nonumber \\
&=& \frac{\partial q}{\partial X} \tan \theta_{canon}\, ,
\label{eq:thetacompgeneral}
\end{eqnarray}
where we used the definition of the momentum (\ref{eq:Pidef}). As discussed before, the Lagrangians we are considering have $\partial q/\partial X \geq 1$, so $\theta_{non-canon} \geq \theta_{canon}$ in Regions B to D in phase space. 
In Region A, however, (\ref{eq:thetacompgeneral}) becomes
\begin{equation}
\tan \theta_{non-canon} = \frac{\partial q}{\partial X} \frac{H_{non-canon}}{H_{canon}} \tan \theta_{canon}
\end{equation}
so the comparison of the angles depends both on the size of $\partial_X q$ and 
$H_{non-canon}$.  In the non-canonical limit, this leads to (\ref{eq:thetacompA}).

\section{Overshoot Trajectories}
\label{sec:overshoot}

Trajectories in Region A are a problem for canonical inflation.  Although trajectories
in this region lose momentum through Hubble friction, as we have seen the angle
the trajectory makes with the horizontal is very small (in Planck units).
Because of this, in canonical inflation, trajectories in Region A typically remain
in Region A, never arriving at the inflationary solution, as in 
Figures \ref{fig:RegionPlots_Inflection} and \ref{fig:RegionPlots_Coulomb}.
(Notice that for the inflection-point potential, even trajectories that start in Region D,
with small momentum, evolve towards Region A.)
These are {\it overshoot trajectories}, and require some degree of fine tuning of the initial
conditions of inflation to avoid \cite{Overshoot}.

\begin{figure}[tp]
\centerline{
\includegraphics[scale=.6]{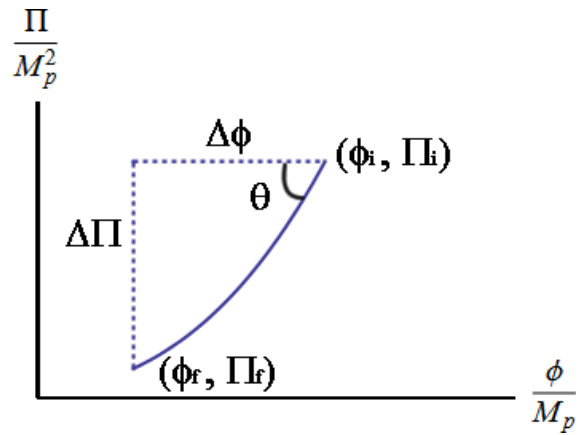}}
\caption{\small A trajectory in phase space makes an angle $\theta$ with the horizontal, and traverses
a certain distance $(\Delta \phi,\Delta \Pi)$.}
\label{fig:AngleDiagram}
\end{figure}

Let us make our discussion of overshoot trajectories somewhat more precise.  Consider a trajectory
that starts at some initial point $(\phi_i,\Pi_i)$ in phase space and ends, some time later,
at $(\phi_f,\Pi_f)$, as in Figure \ref{fig:AngleDiagram}.  For any given
$\{\Pi_i,\Pi_f\}$, the trajectory moves some distance $\Delta \phi$ in field space.
Our primary interest is in estimating the distance the trajectory must travel in order to exit
Region A, as a way to better understand the dynamics of the putative overshoot trajectories.
Denote the typical distance a trajectory must travel in order to exit Region A as $(\Delta \phi)_A$.

One way of gauging the danger that these trajectories pose to the fine tuning of initial conditions
is to compare $(\Delta \phi)_A$ to the size of field space in which inflation occurs $(\Delta \phi)_{inf}$.
We can construct an ``overshoot parameter,"
\begin{equation}
\alpha \equiv \frac{(\Delta \phi)_A}{(\Delta \phi)_{inf}}\, ,
\end{equation}
that measures the relative size of these distances.  Certainly, if a trajectory in Region A only moves
some small fraction of the distance over which inflation occurs, so that $\alpha \ll 1$, then we
do not expect that there will be severe fine-tuning issues.
Alternatively, if $\alpha \gg 1$, then a trajectory moves over a distance much larger than the
size of the inflationary region, and we expect then that fine tuning of initial conditions
will be necessary to avoid these dangerous Region A trajectories.

The parameter $\alpha$ is thus a useful quantity to gauge the severity of large momentum trajectories
for the fine tuning of initial conditions that lead to inflation.  Its advantage over the parameter
$\Sigma$, whose definition as the ratio of the volume of initial conditions that give rise to more than $60$ 
e-folds of inflation to the total volume of phase space is given in Section \ref{subsec:finetuning}, is that $\alpha$ can be computed almost completely
analytically and is still correlated with the more direct measure of fine tuning $\Sigma$.
In addition, as we will see, $(\Delta \phi)_A$ is independent of the details of the potential and is
only sensitive to the structure of the kinetic terms, making it ideal for an analysis in which
non-canonical kinetic terms play an important role.
In contrast, the computation of the fine tuning parameter $\Sigma$ is dependent on the details of the potential.

For a canonical kinetic term, we found that in Region A the angle is
\begin{equation}
\left.\frac{d\Pi}{d\phi}\right|_A = \sqrt{\frac{3}{2}} \frac{\Pi}{M_p}\, .
\nonumber
\end{equation}
This can integrated to obtain
\begin{equation}
\frac{(\Delta \phi)_A^{canon}}{M_p} = \sqrt{\frac{2}{3}} \log \left(\frac{\Pi_i}{\Pi_f}\right)\,.
\label{eq:deltaphicanon}
\end{equation}
The maximum value for the initial momentum is, as discussed before, $\Pi_i \lsim M_p \Lambda$,
while the minimum is set by the lower boundary of Region A, i.e. $\Pi_f \sim \sqrt{2V}$.
However, since the dependence on the initial and final momenta in (\ref{eq:deltaphicanon})
is logarithmically weak, we expect a typical trajectory to move an ${\mathcal O}(1)$ distance
in Planck units before exiting Region A;
\begin{equation}
(\Delta \phi)_A^{canon} \sim M_p\, .
\end{equation}
Notice that this is independent of the type of potential, so it is true for all canonical small-
and large-field models.  

We now see dynamically, then, why small-field models typically have
an overshoot problem but large-field models do not.  For small-field models, the size of the field space in which inflation happens is small in Planck units by definition,
$(\Delta \phi)_{inf} \ll M_p \sim (\Delta \phi)_A$,
so the overshoot parameter is very large: $\alpha \gg 1$.  
This is in agreement with the known fine tuning of initial conditions required in small field models 
\cite{Goldwirth,GoldwirthPiranReport}.
On the other hand, for large-field models we typically have $(\Delta \phi)_{inf} \sim {\mathcal O}(10-100)\times M_p$
so $\alpha \ll 1$.
Trajectories that start with large momentum typically only move over some small fraction of
the size of the inflationary region before losing enough momentum to enter inflation.
Large-field models do not require fine tuning of initial conditions
\cite{Linde,Belinskyetal,PiranWilliams,Piran:1986dh,GoldwirthPiranReport}, in agreement with the size of the
overshoot parameter $\alpha \ll 1$ for these models.

For a non-canonical kinetic term, we can do a similar integration of the angle in Region A,
\begin{equation}
\left.\frac{d\Pi}{d\phi}\right|_A = \frac{\sqrt{3}}{(2R)^{1/4}} \left(\frac{\Pi}{M_p^2}\right)^{3/2} \frac{M_p^2}{\Lambda},
\nonumber
\end{equation}
to obtain
\begin{equation}
\frac{(\Delta \phi)_A^{non-canon}}{M_p} = \frac{(2R)^{1/4}}{2\sqrt{3}} \left[\left(\frac{\Lambda^2}{\Pi_f}\right)^{1/2}
	-\left(\frac{\Lambda^2}{\Pi_i}\right)^{1/2}\right]\, .
\label{eq:deltaphinoncanon}
\end{equation}
Taking $\Pi_i \gg \Pi_f$, we can drop the last term in (\ref{eq:deltaphinoncanon}).  The lower
boundary of Region A occurs for $\Pi/\Lambda^2 \sim (V/\Lambda^4) (2R)^{-1/2}$, so that we have
\begin{equation}
(\Delta \phi)_A^{non-canon} \approx \frac{(2R)^{1/2}}{2\sqrt{3}} M_p \sqrt{\frac{\Lambda^4}{V}}\, .
\end{equation}
We see that if $V/\Lambda^4 \gg 1$, the non-canonical trajectory moves a much smaller distance
than it does with a canonical kinetic term.
In particular, compared to the canonical kinetic term, we have
\begin{equation}
\frac{\alpha_{non-canon}}{\alpha_{canon}} \sim \frac{(2R)^{1/2}}{2\sqrt{3}} \sqrt{\frac{\Lambda^4}{V}} \ll 1\, ,
\end{equation}
so non-canonical kinetic terms can dramatically improve the degree of overshooting.
In the next section, we will see this in more detail by examining a few examples.

\section{Examples}
\label{sec:examples}

Let us now consider some specific examples where the effect of the non-canonical kinetic terms reduces
the overshoot and initial conditions fine-tuning problems.
As mentioned before, we will consider two types of characteristic small-field inflation potentials known
to have overshoot and initial condition fine-tuning problems, the inflection-point (\ref{eq:inflectionpt}) 
and coulomb (\ref{eq:coulomb}) potentials,
\begin{eqnarray}
V(\phi)_{inflection} &=& V_0 + \lambda (\phi-\phi_0) + \beta (\phi-\phi_0)^3\, ; \nonumber \\
V(\phi)_{coulomb} &=& V_0 - \frac{T}{(\phi+\phi_0)^n}\, . \nonumber
\end{eqnarray}
We will choose the parameters below for these potentials such that for a canonical kinetic term, more than 60 e-folds of (slow-roll) inflation leads to an acceptable power
spectrum in the appropriate observational window, consistent with current observations \cite{WMAP}, 
i.e. $P_\zeta = 2.41\times 10^{-9},\ n_s = 0.961$.
\begin{eqnarray}
\mbox{Inflection:}&& 
\begin{pmatrix} 
V_0 = 3.7\times 10^{-16}, & \lambda = 1.13\times 10^{-20} \\
\beta = 1.09\times 10^{-15}, & \phi_0 = 0.01
\end{pmatrix}; \nonumber \\
\mbox{Coulomb:}&& 
\begin{pmatrix} V_0 = 5.35\times 10^{-14}, & T=\phi_0^4 V_0,\\
n=4,& \phi_0 = 0.0
\end{pmatrix}\, .\nonumber
\end{eqnarray}
We will choose the scale $\Lambda$ that controls the strength of the non-canonical kinetic terms to be 
$\Lambda = \{5\times 10^{-6},5\times 10^{-5}\}$ for the inflection-point and coulomb potentials respectively.
These values are chosen so that the background inflationary solution obeys all of the relevant effective field theory
constraints and conditions \cite{EFTInflation,PerturbativeInflation,Shandera:2008ai}, but also still allows the 
non-canonical kinetic terms to play an important role for the dynamics (see \cite{NonCanonAttractors}
for more discussion of these constraints).

As examples, we will compare a canonical Lagrangian to a Lagrangian with a ``geometric series"-type kinetic term:
\begin{eqnarray}
p(X,\phi)_{canon} &=& X-V(\phi) ;\\
p(X,\phi)_{geo} &=& \Lambda^4 \left(\frac{1}{1-X/\Lambda^4}-1\right) - V(\phi) \, .
\label{eq:geo}
\end{eqnarray}
The Lagrangians will be compared for the same potential (inflection point
or coulomb) so that the only thing that differs is their kinetic terms.

\begin{figure*}
\includegraphics[scale=.5]{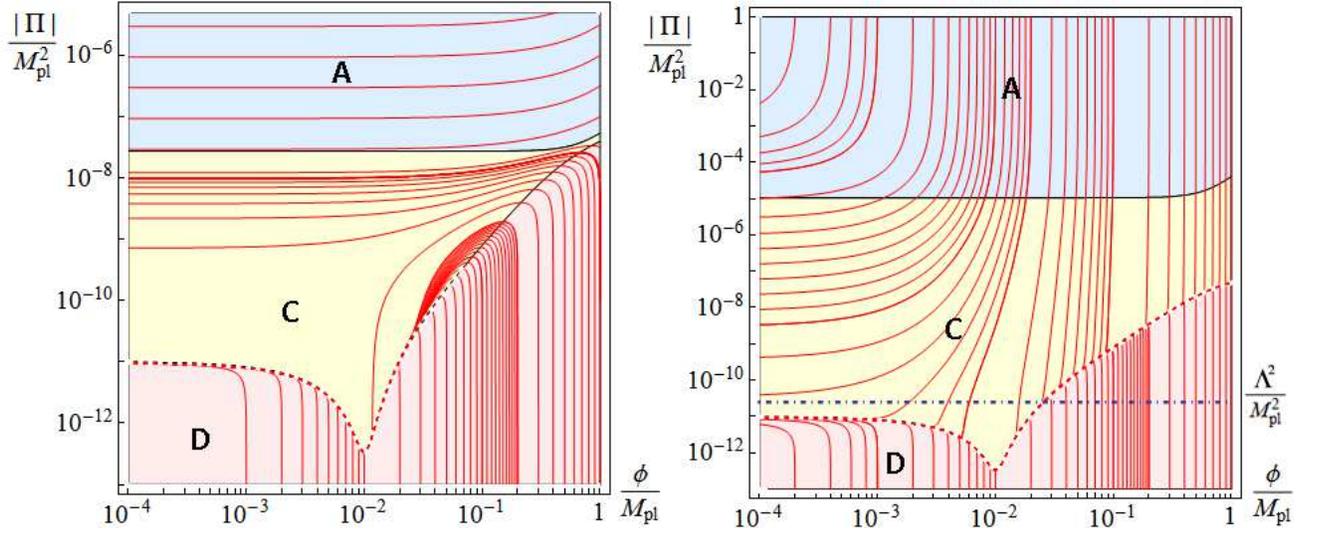}
\caption{\small Left: Phase-space plot for a canonical Lagrangian with an inflection-point potential. Right: Phase-space
plot for the non-canonical Lagrangian (\ref{eq:geo}) with the same inflection-point potential.  Notice that trajectories
that previously were overshoot trajectories for a canonical kinetic term are attracted strongly to the inflationary solution.
The horizontal dot-dashed line is the value $\Pi=\Lambda^2$; above this the system is non-canonical.
}
\label{fig:PhasePlots_Inflection}
\end{figure*}

\begin{figure*}
\includegraphics[scale=.5]{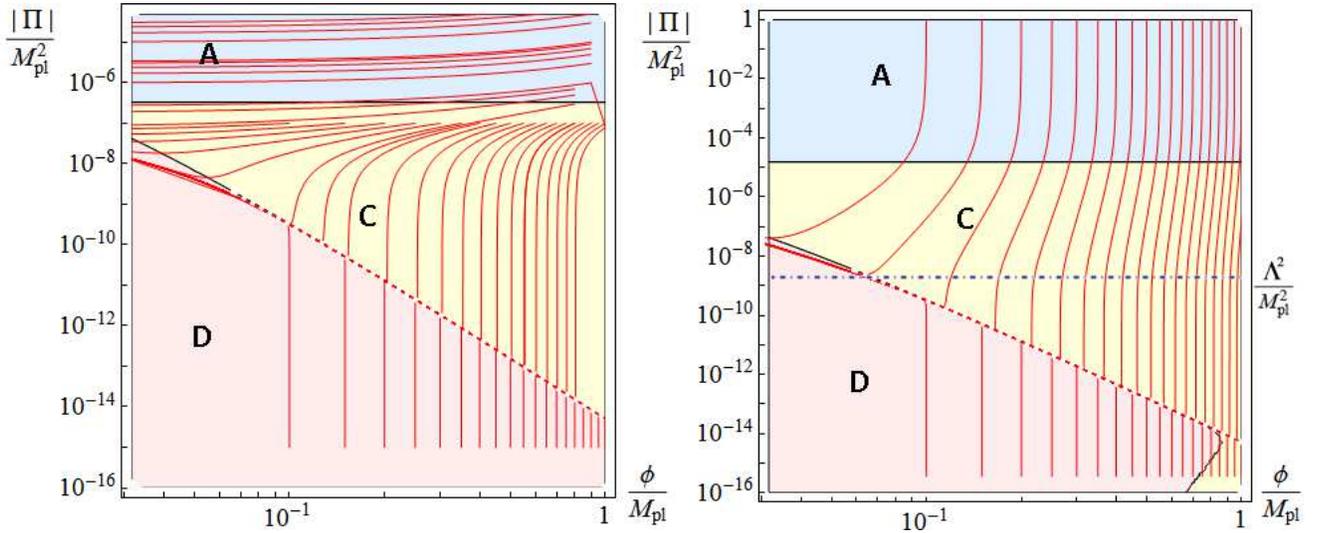}
\caption{\small Left: Phase-space plot for a canonical Lagrangian with a coulomb potential. Right: Phase-space
plot for the non-canonical Lagrangian (\ref{eq:geo}) with the same coulomb potential.  Notice that trajectories
that previously were overshoot trajectories for a canonical kinetic term are attracted strongly to the inflationary solution.
}
\label{fig:PhasePlots_Coulomb}
\end{figure*}

The phase-space plots shown in Figures \ref{fig:PhasePlots_Inflection} and \ref{fig:PhasePlots_Coulomb} 
include the inflationary solution $\Pi_{inf}(\phi)$ (dotted line),
sample trajectories (solid red lines) showing the flow in phase space, and boundaries of the regions
for the inflection-point and coulomb potentials respectively, for both the canonical and non-canonical Lagrangians.
Notice that the qualitative features of the structure and dynamics of phase space discussed throughout Sections 
\ref{sec:structure}-\ref{sec:overshoot} are evident upon comparison of the canonical and non-canonical Lagrangians:
\begin{itemize}
\item The upper boundary of the allowed phase space for $\Pi$ is much larger, by a factor of $M_p/\Lambda$.
\item The lower boundary of Region A is pushed to larger $\Pi$. (Notice that, for these potentials, $\epsilon_{SR} < 3$
in our regime of interest, so Region B does not appear.)
\item The angles that trajectories make with the horizontal are generally much steeper.
\item Overshoot trajectories in Region A are much less prevalent.  In particular, the overshoot
parameter $\alpha \sim \infty$ for the canonical Lagrangian
(trajectories never exit Region A for the small-field regime we are considering), while $\alpha \lsim 10^{-3}$
for the non-canonical Lagrangian, putting it on a par with the overshooting found in typical 
large-field models.
\end{itemize}
From Figures \ref{fig:PhasePlots_Inflection} and \ref{fig:PhasePlots_Coulomb}, we see that non-canonical kinetic
terms modify the dynamics of phase space in such a way as to reduce the prevalence of overshooting.

\subsection{Initial Conditions Fine Tuning}
\label{subsec:finetuning}

It is helpful to make the effects of the non-canonical kinetic terms on the problem of fine tuning of initial conditions
more precise.  Consider a standard measure of the amount of fine tuning of the initial conditions,
the ``initial conditions fine-tuning" parameter, defined as the
fraction of initial-conditions phase space that leads to more than 60 e-folds of inflation,
\begin{equation}
\Sigma = \frac{\int d\phi\, d\Pi\ \theta(N_e-60)}{\int d\phi\, d\Pi}\, ,
\end{equation}
where the number of e-folds is defined as usual, i.e. $N_e \equiv \log\frac{a_e}{a_i} = \int H dt$,
and $\theta(x)$ is $1$ for $x>0$ and $0$ for $x<0$.
This has been used to parameterize the fine tuning of initial conditions before 
\cite{Belinskyetal,PiranWilliams,Goldwirth,Brandenberger:2003py,Birdetal}.

We should make it clear that using $\Sigma$ to quantify the fine tuning of initial conditions presupposes
a flat measure on the $(\phi,\Pi)$ phase space, without any volume weighting, for example.
The choice of a measure for inflation involves many subtleties; see \cite{NaturalMeasure,MeasureCosmo}
for a discussion.
Certainly, for a choice of measure that exponentially disfavors inflation \cite{MeasureCosmo}
it is unlikely that the modified dynamics we find here will significantly change those conclusions.
However, our aim in using $\Sigma$ is not to present it as a definitive measure of the naturalness 
of initial conditions, but rather to investigate how sensitive this simple parameter is to changes
in the kinetic term.  In this way $\Sigma$ serves as a useful diagnostic of the modified dynamics
coming from the non-canonical kinetic term with as few assumptions as possible on the measure of the space
of initial conditions.

\begin{figure*}
\includegraphics[scale=.65]{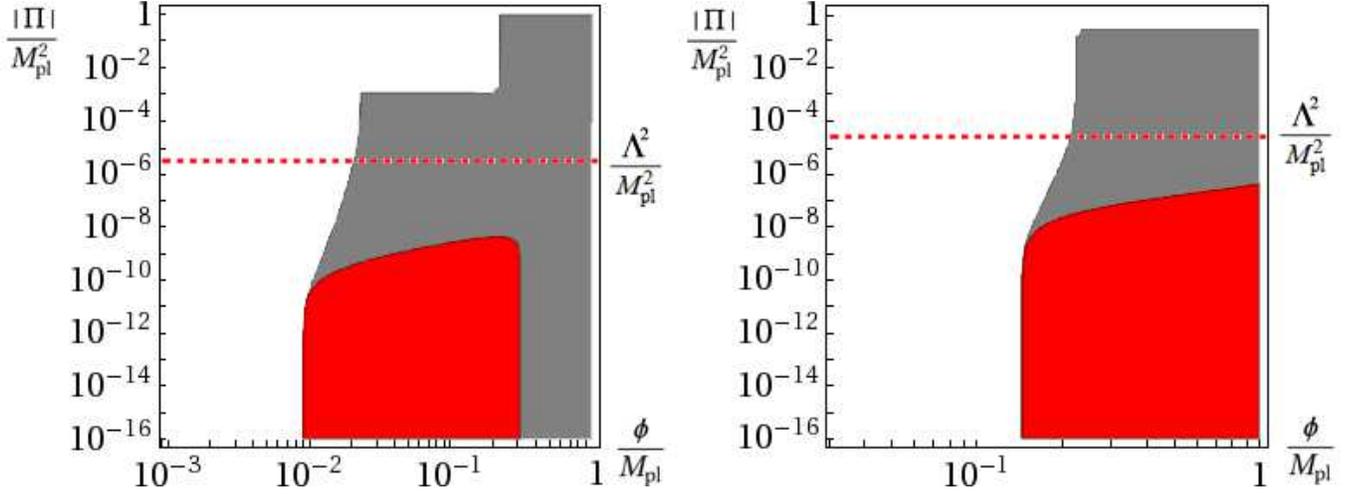}
\caption{\small Left: The shaded regions represent initial conditions that lead to more than 60
e-folds of inflation for an inflection-point potential with canonical kinetic term (red)
and non-canonical kinetic term (gray).  Right: The same, but for a coulomb potential.  Notice
that this is a log plot, so that the size of the 60 e-fold region for the canonical Lagrangian
is exponentially smaller than that for the non-canonical Lagrangian.
The dashed horizontal (red) line denotes the upper boundary of phase space for a canonical
kinetic term (see discussion in Section \ref{sec:structure}).
}
\label{fig:EfoldPlot}
\end{figure*}

In Figure \ref{fig:EfoldPlot} we show the fraction of phase-space initial conditions (in log
scale) that gives rise to more than 60 e-folds of inflation for canonical and non-canonical
kinetic terms and the two different potentials, inflection point (left) and coulomb (right).
We see, as expected, that for a canonical kinetic term (red shaded region) only a very small region of initial-conditions phase space
leads to enough successful inflation, with
\begin{equation}
\Sigma_{canon} \approx \left\{2.4\times 10^{-4},4.5\times 10^{-3}\right\}\, 
\end{equation}
for the inflection-point and coulomb potentials, respectively.
As expected from Figures \ref{fig:PhasePlots_Inflection} and \ref{fig:PhasePlots_Coulomb}, the region
of initial-conditions phase space that leads to enough successful inflation for the non-canonical Lagrangian 
is {\it much} larger for both potentials, with
\begin{equation}
\Sigma_{non-canon} \approx \left\{0.80,0.72\right\}\, 
\end{equation}
(again for inflection point and coulomb potentials respectively).
Thus, the non-canonical Lagrangian effectively does not require any initial-conditions fine tuning at all, in 
contrast to the canonical Lagrangian.

Our results are insensitive to the choice of non-canonical kinetic term.  For example, a DBI Lagrangian, 
\begin{equation}
p(X,\phi) = \Lambda^4 (\sqrt{1-2X/\Lambda^4}-1)-V(\phi),
\end{equation}
gives very similar results to the geometric-series Lagrangian, as shown in Figure \ref{fig:DBI}.
In the Appendix, we extend our results to a ``power-like" class of Lagrangians, finding that they
behave very similarly to the power-series Lagrangians.  Thus, our conclusions do not seem to be
particularly sensitive to the form of the Lagrangian, only whether or not non-canonical kinetic terms
are relevant for inflation.

We see then that non-canonical kinetic terms can drastically reduce the amount of homogeneous initial-conditions fine tuning needed to obtain successful inflation, making these models as ``attractive"
as large-field models.

\begin{figure*}
\includegraphics[scale=.5]{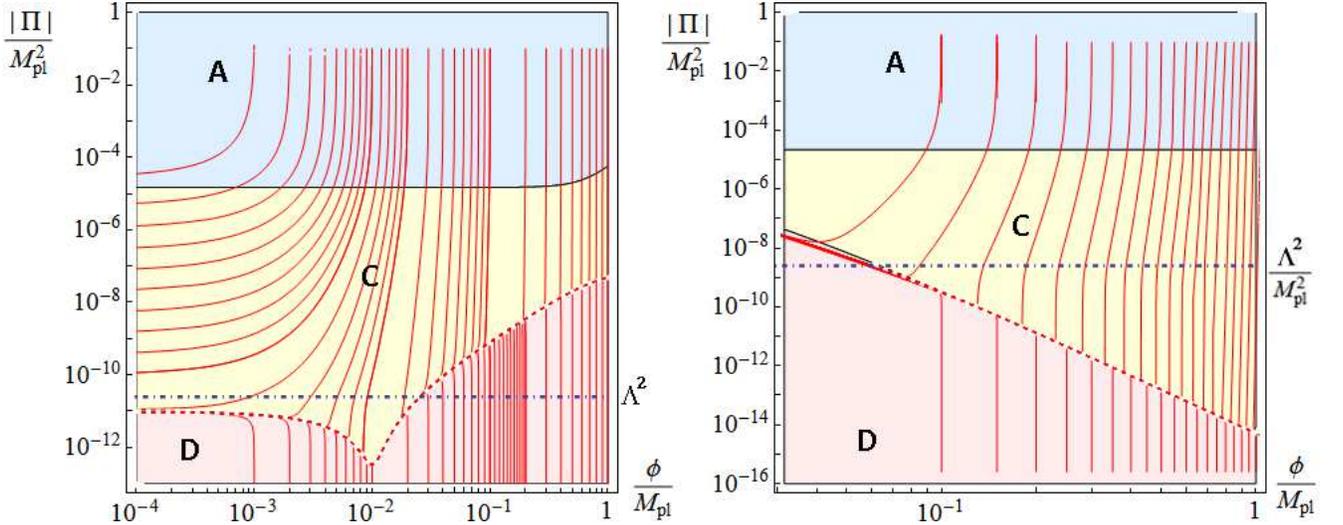}
\caption{\small Left: Phase-space plot for a DBI Lagrangian with an inflection-point potential. Right: Phase-space
plot for a DBI Lagrangian with a coulomb potential.  Notice that the phase-space dynamics are essentially the same as
for the geometric-series Lagrangian in Figures \ref{fig:PhasePlots_Inflection} and \ref{fig:PhasePlots_Coulomb}.}
\label{fig:DBI}
\end{figure*}

\section{Conclusion}
\label{sec:conclusion}

In this paper we investigated the dynamics of (homogeneous) phase space for single-field inflationary
models with both canonical and non-canonical kinetic terms.
Inflationary trajectories are formally attractors in phase space, as discussed in \cite{NonCanonAttractors}, 
but since in practice not all initial conditions 
lead to inflation, some degree of fine tuning is required.
We uncovered general features of the phase space of non-canonical kinetic terms that can be useful
for understanding inflationary initial conditions.

We divided phase space into several regions depending on the dynamics implied by the equation of motion.
Our general regional analysis allowed
us to construct a qualitative picture of the flow of trajectories in phase space, as in 
Figures \ref{fig:RegionPlots_Inflection} and \ref{fig:RegionPlots_Coulomb}.
We investigated the angles that trajectories make in phase space, and found that as the kinetic terms
become more non-canonical, trajectories become steeper.  In particular, trajectories that have a large
amount of momentum lose their momentum through Hubble friction {\it faster} for non-canonical kinetic
terms.  This effect leads to an amelioration of the overshoot problem for inflation, in which a trajectory
with too much kinetic energy never enters the inflationary phase.  

We characterized the prevalence of overshoot trajectories through an ``overshoot parameter" $\alpha$, given by
\begin{equation}
\alpha \equiv \frac{(\Delta \phi)_A}{(\Delta \phi)_{inf}}\nonumber,
\end{equation}
which is the ratio of the field distance a typical trajectory travels in the large-momentum region (Region A).
If $\alpha \ll 1$, then the inflaton only moves a fraction of the size of the inflationary trajectory before
shedding its momentum through Hubble friction, so it does not overshoot.  If $\alpha \gg 1$, the inflaton
must move many times the distance spanned by the inflationary trajectory to shed its momentum through
Hubble friction, so it generically overshoots the inflationary regime.

We computed $(\Delta \phi)_A$ for canonical and non-canonical kinetic terms, finding
\begin{eqnarray}
(\Delta \phi)_A^{canon} &=& \sqrt{\frac{2}{3}} M_p \log\left(\frac{M_p \Lambda}{\sqrt{2V}}\right) \sim {\mathcal O}(1) M_p \nonumber; \\
(\Delta \phi)_A^{non-canon} &=& \frac{(2R)^{1/2}}{2\sqrt{3}} \left(\frac{\Lambda^4}{V}\right)^{1/2} M_p \ll M_p\, .\nonumber
\end{eqnarray}
For a typical canonical small-field inflationary model, $(\Delta \phi)_{inf} \ll M_p$ so $\alpha \gg 1$
and we expect a significant overshoot problem.  In contrast, for large field inflation 
$(\Delta \phi)_{inf} \gsim {\mathcal O}(10-100) M_p$ so $\alpha \ll 1$ and we expect no overshoot problem.
These qualitative results are in agreement with the numerical results 
\cite{Goldwirth,GoldwirthPiranReport,Belinskyetal,PiranWilliams,Piran:1986dh}.
Importantly, the distance traveled while shedding momentum through
Hubble friction is significantly smaller for non-canonical kinetic terms.  For specific examples, we found the non-canonical kinetic
terms decreased the value of $\alpha$ from $\alpha \sim \infty$ for canonical small-field models 
(trajectories never leave Region A) to $\alpha \sim 10^{-3}$
for non-canonical models.
Correspondingly, the prevalence of overshoot trajectories is significantly decreased, as can be
seen in Figures \ref{fig:PhasePlots_Inflection} and \ref{fig:PhasePlots_Coulomb}.

We also investigated the effect of the non-canonical kinetic terms on the standard 
initial-conditions fine-tuning problem.  Because the non-canonical kinetic terms 
significantly modify the dynamics in the large-momentum region, we found that small-field 
models with non-canonical kinetic terms do not have an initial-conditions fine-tuning problem, 
even when their canonical counterparts do.  In particular, in terms of
the fraction $\Sigma$ of phase space leading to $60$ e-folds of inflation,
we studied small-field inflationary potentials which required
$\Sigma \sim 10^{-4}-10^{-3}$ for canonical kinetic terms, but $\Sigma \sim 0.8$ for non-canonical kinetic terms.

Of course, not every effective theory for inflation with non-canonical kinetic terms leads to such drastic improvements
in the overshoot and initial-conditions fine-tuning problems.  Certainly, if the energy $\Lambda$ that controls
the scale at which the corrections to the kinetic term become important is too large, we do not expect
to see a significant effect.  In our analysis, we assumed that some period of non-canonical inflation occurs;
this appears to be sufficient to affect the dynamics in phase space, but it is not clear if it is necessary.

It is difficult to make progress on a theory of inflationary initial conditions because we have so little
information about the initial conditions: by construction inflation washes out features of the pre-inflationary era.
However, the dynamics of the inflaton, visible through cosmological observables such as
the tensor-to-scalar ratio or equilateral non-gaussianity 
may give important clues as to the pre-inflationary physics.
Through the (generalized) Lyth bound \cite{Lyth,BaumannMcAllister},
\begin{equation}
\frac{\Delta \phi}{M_p} = \int_0^{N_{end}} \sqrt{\frac{r}{8}} \frac{dN_e}{\sqrt{c_s P_X}} 
	\approx \frac{{\mathcal O}(1)}{\sqrt{c_sP_X}} \left(\frac{r}{0.01}\right)^{1/2}\, , \nonumber
\end{equation}
we learn that observation of a tensor-to-scalar ratio $r \gsim 0.01$ would imply that inflation is a large field
model, and so the standard dynamics of a canonical kinetic term would be enough to drive the (homogeneous)
pre-inflationary universe towards inflation.  Similarly, because the size of primordial equilateral non-gaussianity
depends on the deviation of the system from a canonical kinetic term \cite{NonGauss}, via
\begin{equation}
f_{NL}^{(equil)} \sim c_s^{-2}\, , \nonumber 
\end{equation}
an observation of equilateral non-gaussianity would suggest that the dynamics of the non-canonical kinetic
terms are important for driving the pre-inflationary universe towards inflation. 
It is possible that there are other observable features that may help us refine our
ideas about the physics of inflationary initial conditions \cite{Freivogel:2005vv,Itzhaki2}.

Certainly, a more robust treatment of inflationary initial conditions must include the effect of 
inhomogeneities.  It is not clear if, or how, non-canonical kinetic terms modify the dynamics of
inhomogeneities in the pre-inflationary universe, questions which would be interesting to investigate further.

\section*{Acknowledgments} We would like to thank
Daniel Baumann,
Robert Brandenberger, Cliff Burgess, Jim Cline, Louis Leblond, Liam McAllister, Enrico Pajer, Hiranya Peiris, Sarah Shandera,
and Henry Tye for helpful discussions.  
The work of P.F.~is supported by the Natural Sciences and Engineering Research Council (NSERC) of Canada. 
R.G.~is supported by an NSERC Postdoctoral Fellowship, and the STRFC grant ST/G000476/1 ``Branes, 
Strings and Defects in Cosmology," as well as by a Canada-UK Millennium Research Award.
B.U.~is supported in part through an IPP (Institute of Particle Physics, Canada) Postdoctoral Fellowship, and
by a Lorne Trottier Fellowship at McGill University and would like to thank the 
Aspen Center for Physics for hospitality while part of this work was completed.
The work of A.W.~is supported by the Fonds Qu\'eb\'ecois de la Recherche sur la Nature et les Technologies (FQRNT).

\appendix

\section{Power-like Lagrangians}

In the text we considered only the ``power-series" type of Lagrangian, namely Lagrangians which lead to non-canonical
inflation when expressed as a power series within its domain of convergence.
As was discussed in \cite{NonCanonAttractors}, there is another class of Lagrangians that give rise
to non-canonical inflation {\it outside} of their domain of convergence when expressed as a power series.
It is not clear whether these Lagrangians make sense as effective field theories (EFTs), since we would expect
the validity of the EFT to break down at the radius of convergence of the power series.
Nevertheless, we consider these Lagrangians in this appendix for completeness, in case they can
be seen as sensible EFTs.

The typical example from this class is the ``power-like" Lagrangian
\begin{equation}
p(X,\phi) = \Lambda^4 \left[\left(1+\frac{1}{n}\frac{X}{\Lambda^4}\right)^n-1\right]-V(\phi)\, .
\label{eq:powerlike}
\end{equation}
For this Lagrangian, the canonical momentum is
\begin{equation}
\Pi = \sqrt{2X} \left(1+\frac{1}{n}\frac{X}{\Lambda^4}\right)^{n-1}\, .
\end{equation}
In the non-canonical regime, we see that $X/\Lambda^4 \gg 1$ when $\Pi/\Lambda^2 \gg 1$, so $\Pi\sim \Lambda^2$ is
the boundary of the non-canonical regime in field space as before.
We will restrict the rest of this analysis to this regime, since this is where the phase space will differ from
the canonical case.

As before, the boundaries of the Regions A-D are determined by which term in (\ref{eq:PiEOM},\ref{eq:HEOM})
dominates.  The kinetic energy dominates over the potential energy when
\begin{eqnarray}
\Pi > && \frac{V^{(n-1/2)/n}}{\Lambda^{2(1-1/n)} (2n)^{(n-1)/2n} } \nonumber \\
	&& \stackrel{n\gg 1}{\rightarrow} \frac{V(\phi)}{\sqrt{2n}\Lambda^2},
\end{eqnarray}
which becomes the same as (\ref{eq:noncanonKE}) in the large $n$ limit.
Similarly, the Hubble friction domination condition, when the energy density is mostly kinetic, becomes
(in the large $n$ limit)
\begin{equation}
\Pi > \left(\frac{\epsilon_{SR}^2}{18}\right)^{1/6} \frac{\sqrt{2V}}{(2n)^{1/6}} \left(\frac{V}{\Lambda^4}\right)^{1/6},
\end{equation}
which is the same as (\ref{eq:noncanonHubbleKE}).
The angles in these regions are also modified:
\begin{equation}
\tan \theta = \begin{cases}
\frac{\sqrt{3}}{(2n)^{(n-1)/2(2n-1)}} \frac{\Pi^{(3n-2)/(2n-1)}}{M_p^2\Lambda^{(2(n-1)/(2n-1)}} ;& \cr
-\frac{\sqrt{2\epsilon_{SR}}}{(2n)^{(n-1)/(2n-1)}}\frac{V}{M_p^2 \Lambda^{4(n-1)/(2n-1)} \Pi^{1/(2n-1)}} ;& \cr
\frac{\sqrt{3V}}{(2n)^{(n-1)/(2n-1)}}\frac{\Pi^{(2n-2)/(2n-1)}}{M_p^2\Lambda^{4(n-1)/(2n-1)}}\, . & \cr
\end{cases}
\end{equation}
In the large $n$ limit these become
\begin{equation}
\tan \theta \stackrel{n\gg 1}{\rightarrow} \begin{cases}
\frac{\sqrt{3}}{(2n)^{1/4}} \frac{\Pi^{3/2}}{\Lambda M_p^2}& \mbox{(A)}; \cr
-\frac{\sqrt{2\epsilon_{SR}} V}{\sqrt{2n} M_p^2 \Lambda^2} & \mbox{(B) and (D)}; \cr
\frac{\sqrt{3 V} \Pi}{\sqrt{2n} M_p^2 \Lambda^2} & \mbox{(C)}\, . \cr
\end{cases}
\end{equation}
We see that these also approach the same form as the power-series Lagrangians in the large $n$ limit.
Thus these Lagrangians give very similar results.

\bibliography{kAttractors}

\end{document}